# Quantifying indirect and direct vaccination effects arising in the SIR model

Lixin Lin[1], Homayoun Hamedmoghadam[2], Robert Shorten[2], Lewi Stone[1,3*]

[1]Mathematical Sciences, School of Science, RMIT University, Melbourne, Australia
[2]Dyson School of Design Engineering, Imperial College London, London, UK
[3]Biomathematics Unit, School of Zoology, Faculty of Life Sciences, Tel Aviv University, Tel Aviv, Israel

**Abstract**

Vaccination campaigns have both direct and indirect effects that act to control an infectious disease as it spreads through a population. Indirect effects arise when vaccinated individuals block disease transmission in any infection chains they are part of, and this in turn can benefit both vaccinated and unvaccinated individuals. Indirect effects are difficult to quantify in practice, but here, working with the Susceptible-Infected-Recovered (SIR) model, they are analytically calculated in important cases, through pivoting on the Final Size formula for epidemics. Their relationship to herd immunity is also clarified. Furthermore, we identify the important distinction between quantifying indirect effects of vaccination at the "population level" versus the "per capita" individual level, which often results in radically different conclusions. As an important example, the analysis unpacks why population-level indirect effect can appear significantly larger than its per capita analogue. In addition, we consider a recently proposed epidemiological non-pharmaceutical intervention used over COVID-19, referred to as "shielding", and study its impact in our mathematical analysis. The shielding scheme is extended by inclusion of limited vaccination.

## 1. Introduction

Vaccination campaigns have both direct and indirect effects on the transmission of an infectious disease as it spreads through a population [1-4]. Direct vaccination effects refer to the reduction in the risk of infection due to the protection provided to individuals by the vaccine dose. Indirect vaccination effects refer to the protection provided to an individual by any nearby vaccinated neighbors that indirectly act to block chains of incoming infections, or just reduce infection possibilities, thereby "shielding" the individual [3-5]. The indirect effect of vaccination, via such shielding, makes it more difficult for a disease to spread in a population and hence also affects conditions that give rise to herd immunity [6]. Indirect vaccination effects are difficult to quantify in practice, but clearly understanding the role and magnitude of indirect effects is critical for assisting in the design of vaccination programs [2-4]. A deeper theoretical analysis is still lacking and there is a need to better formulate conditions that predict when the impact of indirect effects will be significant and when they will be minor [2-4, 7]. Here we make use of simple mathematical models to achieve this goal, and in important cases succeed to give an exact characterization of indirect effects. In the process, we make clear the important distinction of quantifying indirect effects either at the population level or as a "per-capita" quantity, a key distinction that is often neglected.

    Haber [3] gives a simple elegant example that shows how indirect effects matter and how their "shielding" property can be taken advantage of. "Consider a population that consists of 10,000

---

individuals in 2,000 households so that the average household size is 5. If 4,000 vaccine doses are available, what is the best way to distribute the vaccine? Perhaps, the simplest way is to select 800 households and vaccinate everyone in these households. However, an alternative plan, is to vaccinate two persons in each of the 2,000 households. The second plan is more effective, because it utilizes the indirect effects of shielding to protect the unvaccinated household members." In fact indirect effects endow additional protection to all households and individuals in the second plan, in contrast to the first plan that leaves 1,200 households with no protection.

The most striking example of indirect effects relates to herd immunity, where the theory predicts that there is no need to vaccinate more than a critical proportion of a population to completely protect the whole population from a disease [6, 8, 9]. The threshold is predicted, for example, by the simple classical SIR epidemic model we will be describing shortly. At the herd immunity threshold, if a proportion $v$ of the population is "directly" vaccinated, then the proportion $(1-v)$ of the population does not require vaccination since it will be protected indirectly by herd immunity effects [6, 8]. The division of the population in this ratio $v/(1-v)$ (i.e., direct : indirect) proves to be important in what follows.

As mentioned, quantifying the direct and indirect effects is of great value in the evaluation of vaccination campaigns [1-4, 10]. But it is a complicated procedure. In a number of studies, the indirect vaccination effect has been estimated to be unusually large in magnitude. Scutt et al. [11] showed that in some cases the indirect vaccination effect can be more than 400% of the direct vaccination effect. Eichner et al. [4] discussed an example in Canada, where "vaccination of 83% of children (≤15 years) reduced influenza infection incidence in unvaccinated individuals by 61%." Similarly, in modelling COVID-19, Gavish et al. [12] found that among the cases reduced as a result of the booster campaign, ~54% were reduced because of direct protection, whereas the remainder were reduced by indirect protection. Gallagher et al. [13] emphasized the critical importance of considering the indirect vaccination effect when evaluating SARS-CoV-2 vaccine candidates. Through modelling, they emphasized the importance of not automatically selecting a vaccine based solely on the largest direct effect. Weidemann et al. [7] pointed out that there are exceptions and that some studies report that the impact of indirect effects can be low (e.g., as in the school vaccination programs on US county level [14, 15]).

Mathematical models are used here to explore the extent of indirect effects and the factors that enhance them. In contrast to other analyses [4, 11], we use the Final Size formula [16] of the epidemic which allows us to draw analytical conclusions from the mathematical models without relying on numerical simulations.

## 2. SIR epidemics with vaccination

The standard SIR model assumes that at any time $t$, each individual in a population can only belong to any one of three classes: **S**usceptible, **I**nfected or **R**ecovered. Taking the proportions of individuals as $S$, $I$ or $R$, then clearly $S(t) + I(t) + R(t) = 1$. In a randomly mixing population, new infections are generated when infected individuals come into contact with susceptible individuals at a rate proportional to the product $S \cdot I$, which leads to the following well known SIR equations:

$$\frac{dS}{dt} = -\beta SI,$$

$$\frac{dI}{dt} = \beta SI - \gamma I, \tag{2.1}$$
$$\frac{dR}{dt} = \gamma I.$$

In this scheme, over time individuals move from the $S$ class to the $I$ class, and finally end up in the recovered $R$ class i.e., $S \to I \to R$. Here $\beta$ is the transmission rate ($S \to I$) between individuals, while $\gamma$ is the recovery rate ($I \to R$). The basic reproduction number is defined as $R_0 = \beta/\gamma$. Initial conditions are such that almost all members of the population are susceptible, and the number of initially infected individuals are infinitesimally small, so that we may approximate $S(0) = 1, I(0) = R(0) = 0$. According to standard theory, for these initial conditions, if a single infected individual enters the system, an epidemic will occur only if $R_0 > 1$, since it ensures that $dI/dt > 0$ at $t = 0$.

A simple model of vaccination can be explored by changing the initial conditions. Suppose that initially a proportion $v$ of a fully susceptible population has been vaccinated and has thus become fully protected from infection. The population's vaccination coverage is said to be $v$, and we set:

$$S(0) = 1 - v, \quad I(0) = 0, \quad R(0) = v. \tag{2.2}$$

In this case, standard theory shows that an epidemic can only occur if the effective reproduction number $R_e = R_0 \, S(0) > 1$, which is equivalent to vaccination level $v < v_h$ where the herd immunity threshold is:

$$v_h = 1 - \frac{1}{R_0} \tag{2.3}$$

Under this condition, the epidemic's Final Size or proportion of the population infected over the epidemic, is given by $Z(v)$ and is the solution of the following equation [16]:

$$Z(v) = (1 - v)\left(1 - e^{-R_0 Z(v)}\right). \tag{2.4}$$

The solid lines in Fig. 1, show the Final Size $Z(v)$ as a function of vaccination $v$, repeated for different values of the reproduction number $R_0$. In the absence of vaccination ($v = 0$), we denote the Final Size as $Z^* = Z(0)$ and is given by the solution of:

$$Z^* = 1 - e^{-R_0 Z^*}. \tag{2.5}$$

The Final Size $Z^*$ in the absence of vaccination is plotted in Fig. 1a (dashed lines), as a reference indicating the case where $v = 0$. Thus, for a high reproduction number (e.g., $R_0 = 4$, yellow curve), and no vaccination ($v = 0$), the Final Size $Z^* = 0.99$ and the epidemic would infect 99% of the population. The Final Size decreases with $v$, and for any $R_0$ is zero at the herd immunity threshold $v = v_h = 1 - 1/R_0$. Given $R_0 = 4$, at the herd immunity threshold ($v = 0.75$), an epidemic is not possible despite the fact that 25% of the population remain unvaccinated and thus susceptible. In this situation, the vaccination is protecting the unvaccinated population indirectly.

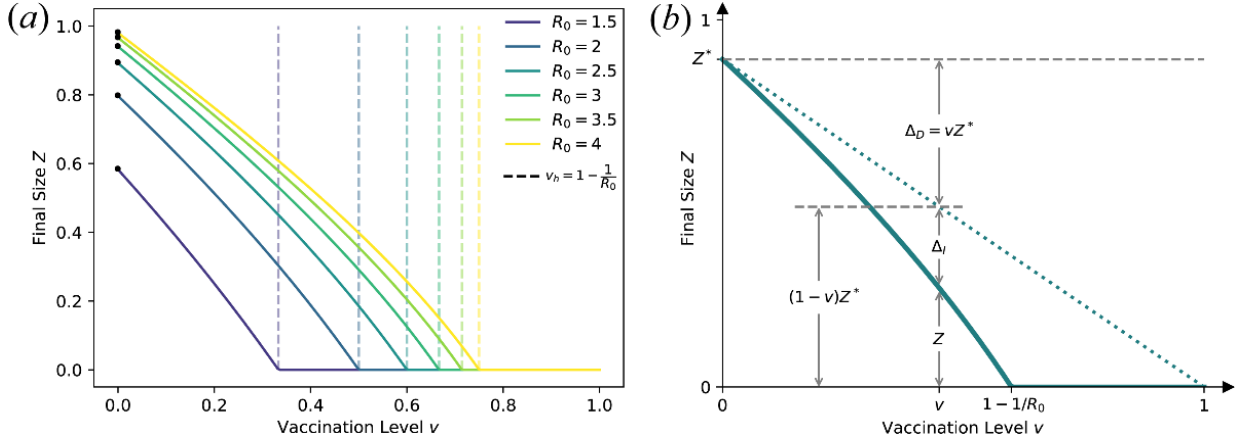

**Figure 1.** (*a*) The relationship between the Final Size of the epidemic and the proportion $v$ of individuals vaccinated for $R_0 = 1.5, 2, 2.5, 3, 3.5, 4$ as calculated by solving Eq. (2.4). Note that the herd immunity threshold, e.g., for $R_0 = 2.5$ is at $v = 0.6$. The black marked points identify the Final Size of the epidemic when $v = 0$, i.e., $Z^*$ for different values of $R_0$. (*b*) The relationship between the proportion of individuals vaccinated, $v$, and the infections averted by direct vaccination effect $\Delta_D = vZ^*$, and indirect vaccination effect $\Delta_I = (1-v)Z^* - Z$. See Eq.2.8 and 2.9.

## 2.1. Calculating direct and indirect effects of vaccination

For any level of vaccination coverage $v$, it is possible to determine the total proportion of the population protected by the vaccine or equivalently the proportion $\Delta_T$ of infections that has been averted, as compared to the situation had there been no vaccination. Most methods use simulation or approximation techniques to achieve this [4, 11, 12]. Here we find a solution based on the Final Size equation for the SIR model. The proportion $\Delta_T$ is simply the difference between the Final Size with no vaccination $Z^*$, and the Final Size with vaccination $Z = Z(v)$, namely:

$$\Delta_T = Z^* - Z. \tag{2.6}$$

The proportion of total infections averted, $\Delta_T$, as a function of $v$ can be visualized as the gap between the solid line for $Z$ and the dashed line for $Z^*$ in Fig. 1a. It is important to note that $\Delta_T$ has two components:

  i) the direct vaccination effect ($\Delta_D$), which is the proportion of infections averted among those vaccinated who achieved a level of protection provided by the vaccine,
  ii) and the indirect vaccination effect ($\Delta_I$), which is the proportion of infections averted among the unvaccinated individuals due to population-level immunity.

That is:

$$\Delta_T = \Delta_D + \Delta_I. \tag{2.7}$$

It is not straightforward to disentangle the two components $\Delta_D$ and $\Delta_I$. In the case of the SIR model, since there is no Vaccinated compartment, it is difficult to track the fate of vaccinees and thus obtain the direct effect $\Delta_D$. Eichner [4] describes a method that adds a Vaccinated compartment and simulates the number of infections generated by vaccinees who received a completely ineffective vaccine. Tallying the total number of new infections generated from this subgroup gives the direct effect. Based on a similar concept we show, using a Final Size formulation, that the proportion of infections directly averted $\Delta_D$ is given by:

$$\Delta_D = v \cdot Z^*. \tag{2.8}$$

where $Z^* = Z(0)$, is the size of the epidemic in the absence of vaccination. Supplementary Note 1 provides a proof based on the continuous-time SIR model.

Equation (2.8) is an important result that is also to some degree intuitive. For example, if the epidemic infected all members of the population ($Z^* = 1$), it is clear that the proportion of those infected who are vaccinated is: $\Delta_D = vZ^*$. The relationship in Eq. (2.8) was also noted by Scutt et al. [11] for a different discrete time epidemic model.

From Eqs. (2.6-2.8), the proportion of infections averted by indirect vaccination as a function of $v$ is:

$$\Delta_I = \Delta_T - \Delta_D = Z^* - Z - v Z^* = (1 - v)Z^* - Z \leq (1 - v)Z^*, \tag{2.9}$$

where again we use the notation that $Z = Z(v)$. The indirect effect thus comprises all unvaccinated infected individuals $(1 - v)Z^*$ that there would have been with no vaccination, but taking away the total number of infected individuals under a vaccination program (see Fig. 1b).

Notice in Fig. 1a that the Final Size as a function of $v$ may be roughly approximated by the linear relationship $Z = Z^*(1 - v/(1 - 1/R_0))$ as long as $v \leq v_h$. The approximation ensures that $Z(0) = Z^*$, and at the herd immunity threshold $v_h = 1 - 1/R_0$, it gives $Z(v_h) = 0$. Substituting this linear expression into Eq. (2.9) results in:

$$\Delta_I \simeq Z^* - Z^*\left(1 - \frac{v}{1 - \frac{1}{R_0}}\right) - v Z^* = \frac{v Z^*}{R_0 - 1}. \tag{2.10}$$

The proportions of infections averted by direct ($\Delta_D$) and indirect ($\Delta_I$) vaccination effect respectively, are shown in Fig. 1b and found from solving Eqs. (2.4-2.9). In Fig. 1b, the dotted green line shows the linear relationship between $(1 - v)Z^*$ and $v$. Thus, the infections averted by direct vaccination effect is the gap between the dashed and the dotted lines, and the infections averted by indirect vaccination effect $\Delta_I$ is the gap between the dotted and the solid green lines.

Figure 2 unpacks the impact of the vaccination effect as a function of $v$. The infections averted by total, direct and indirect vaccination effect respectively are plotted in Fig. 2 as a function of $v$, for $R_0$=1.5, 2.0, 2.5, 3.0, 3.5, 4.0. In the Supplementary Note 2 we use an SIRV simulation model with a vaccination compartment to independently determine $\Delta_T$, $\Delta_D$, and $\Delta_I$ and retrieve identical results that corroborate those shown in Fig. 2.

There are several interesting features seen in Fig. 2. First, the total infections averted by vaccination, $\Delta_T$ clearly increases with $v$, until the herd immunity threshold is reached at $v_h = 1 - 1/R_0$. Hence the most effective campaign is the one with highest vaccination levels despite the presence of prominent indirect effect at lower vaccination levels. Second, the proportion of infections averted by the direct vaccination effect $\Delta_D$ is linear in $v$. In contrast, the proportion of infections averted by the indirect vaccination effect $\Delta_I$ increases nonlinearly in $v$ as long as $v < v_h = 1 - 1/R_0$. Beyond the herd immunity threshold ($v > v_h$), the indirect effect decreases linearly until $\Delta_I = 0$ is reached when $v = 1$. When $v$ is increased beyond the herd immunity threshold $v_h$, the epidemic has died out and there are zero infectives in the population.

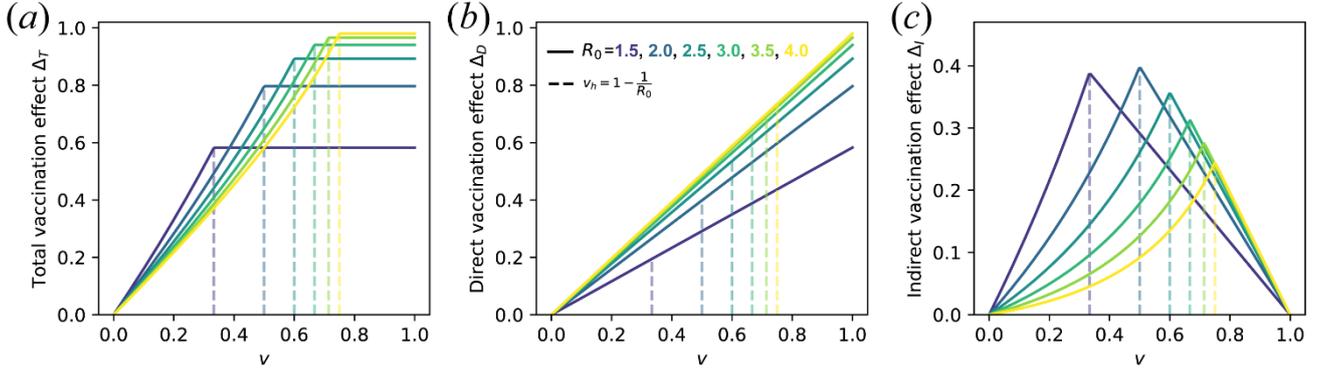

**Figure 2.** The proportion of infections averted by vaccination as a function of vaccination level $v$. (*a*) Total vaccination effect $\Delta_T$; (*b*) Direct vaccination effect $\Delta_D$; (*c*) Indirect vaccination effect $\Delta_I$. This is repeated for $R_0 = 1.5, 2.0, 2.5, 3.0, 3.5, 4.0$. The dashed lines in each panel represent the "herd immunity" vaccination coverage levels $v_h = (1 - 1/R_0)$, with values from left to right corresponding to increasing $R_0$ values.

## 2.2. Ratio of indirect to direct vaccination effects

According to the formulae developed so far, the ratio of indirect to direct vaccination effects is:

$$\frac{\Delta_I}{\Delta_D} = \frac{(1-v)Z^* - Z}{v\, Z^*} = \frac{1-v}{v} - \frac{Z}{v\, Z^*}. \tag{2.11}$$

In Fig. 3a, the ratio $\Delta_I/\Delta_D$ is plotted for different $R_0$ values. Three properties of this ratio are evident from Eq. (2.11) and making use of the fact that $Z = 0$ when $v \geq v_h$ for any specified $R_0$:

i) when $v \geq v_h$, the ratio $\Delta_I/\Delta_D$ can be determined by:

$$\frac{\Delta_I}{\Delta_D} = \frac{(1-v)Z^* - Z}{v\, Z^*} = \frac{1-v}{v}; \tag{2.12}$$

ii) when $v \geq v_h$, the ratio $\Delta_I/\Delta_D$ reaches its maximum at $v = v_h$., we get:

$$\left.\frac{\Delta_I}{\Delta_D}\right|_{(v=v_h)} = \frac{1-v_h}{v_h} = \frac{1}{R_0 - 1}; \tag{2.13}$$

iii) the ratio reaches zero when the entire population is vaccinated:

$$\left.\frac{\Delta_I}{\Delta_D}\right|_{(v=1)} = \lim_{v \to 1}\frac{1-v}{v} = 0; \tag{2.14}$$

In Supplementary Note 3 we show that the indirect- to direct-effect ratio, when only a few individuals are vaccinated (i.e., $v = 0^+$), can be calculated analytically as:

$$\left.\frac{\Delta_I}{\Delta_D}\right|_{(v=0^+)} = \frac{1}{1 - R_0(1 - Z^*)} - 1. \tag{2.15}$$

Recall that from Eq. (2.10), as long as $v \leq v_h$, $\Delta_I \simeq v\, Z^*/(R_0 - 1)$. Thus, a simple approximation for the ratio $\Delta_I/\Delta_D$ in the outbreak phase can be obtained by:

$$\frac{\Delta_I}{\Delta_D} \simeq \frac{v\, Z^*}{(R_0 - 1)(v\, Z^*)} = \frac{1}{R_0 - 1}. \tag{2.16}$$

This approximation is confirmed in Fig. 3b where the exact ratio $\Delta_I/\Delta_D$ is compared with the RHS of Eq. (2.16) (dotted grey curve), both versus $R_0$. Figure 3b shows the ratio $\Delta_I/\Delta_D$ as a function of the basic reproduction number $R_0$, for $v = 0.1, 0.2, 0.3, 0.4$, in all cases for a range of $R_0$ where the vaccination level falls below the herd immunity threshold, i.e., $R_0 > 1/(1-v)$.

The approximation in Eq. (2.16) shows that the indirect and direct effects of vaccination are approximately equal when $R_0 = 2$. In the regime of $1 < R_0 < 2$, the indirect vaccination effect can be significantly larger than the direct vaccination effect. For $R_0 > 2$, the indirect vaccination effect is always less than the direct vaccination effect, and it is almost negligible for $R_0 > 4$. The dotted grey line shows the relationship between the ratio and $R_0$ as predicted by Eq. (2.16). The black marked points in Figs. 3b and Fig. 3c identify where $v=1-1/R_0$, indicating the respective lowest $R_0$ required to initiate an epidemic given a vaccination level.

In a related study, Eichner et al. [4] also examined the ratio of indirect to direct vaccination effects in an SIR model having a vaccination compartment. However, their analysis assumes the presence of demographic birth and death processes, and is carried out when the system reaches an endemic equilibrium state, rather than over a dynamic epidemic as done here. Nevertheless, and intriguingly, they still find the ratio is given by Eq. (2.16) namely $\Delta_I/\Delta_D = 1/(R_0 - 1)$. Using an approximate form of the number of infected individuals for a discrete-time SIR model and examining the mortality rate of the disease, Scutt et al. [11] calculated the ratio $\Delta_I/\Delta_D$ for deaths averted.

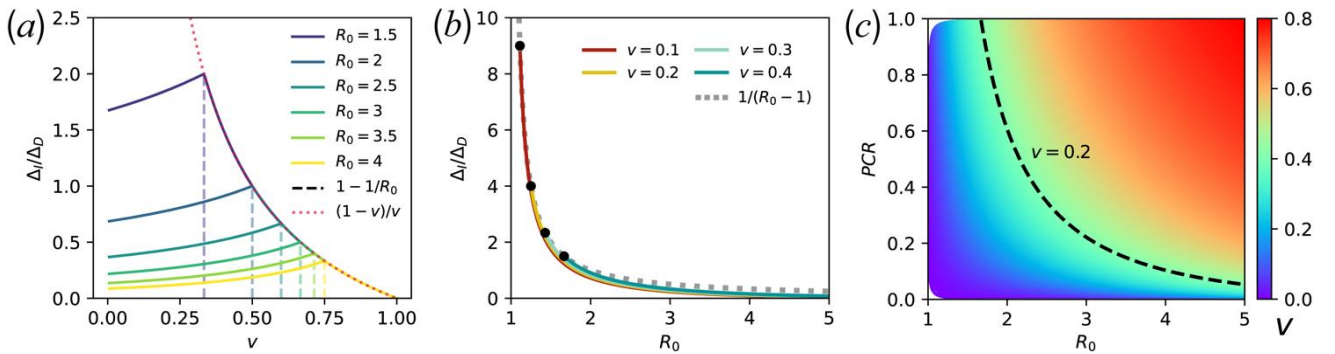

**Figure 3.** (*a*) The ratio of indirect to direct vaccination effects $\Delta_I/\Delta_D$ versus the vaccination level $v$, for different values of $R_0$; using Eq. (2.11). The dashed lines (each associated with the curve of the same color) indicate where herd immunity has been reached for a particular value of $R_0$. The dotted red curve shows $(1-v)/v$, which the ratio collapses to, when $v \geq v_h$ for any value of $R_0$ (see Eq. (2.12)). (*b*) The ratio of indirect to direct vaccination effects $\Delta_I/\Delta_D$ as a function of $R_0$ for different values of $v$. The dashed grey line plots $1/(R_0 - 1)$; see Eq. (2.16). (*c*) PCR as a function of $R_0$ for different vaccination levels $v$ between 0 and 0.8; PCR (Eq. (2.17)) is the indirect vaccination effects per unvaccinated individual against the direct vaccination effects per vaccinated individual. The dashed black curve shows the trend in PCR versus increasing $R_0$ for $v = 0.2$.

## 2.3. When does the indirect vaccination effect play a major role?

The indirect vaccination effect can be substantial, in some cases exceeding 900% of the direct vaccination effect as seen in Fig. 3b. However, here, we show that this does not imply that unvaccinated individuals gain more benefit from the vaccination campaign than vaccinated individuals. To this end, we use the theoretical analysis presented so far and an intriguing example inspired by [11].

Consider a population of $N = 500{,}000$ where 50,000 susceptible individuals were vaccinated ($v = 0.1$). A virus with basic reproduction number $R_0 = 1.2$ invades the population. By solving Eq. (2.4) we obtain that in the ensuing epidemic, $N \cdot Z(0.1) = 65{,}000$ individuals become infected. Had there been no vaccine, from Eq. (2.5) we find the estimate indicating that $NZ^* = NZ(0) = 157{,}000$ would have been infected altogether. Thus, because of the vaccine $N(Z^* - Z(v)) = 92{,}000$ people avoided infection. This can be broken down into:

i) direct vaccination effect, i.e., $N\Delta_D = v\,N\,Z^* = 15{,}700$ infections were averted in the vaccinated population, as found from Eq. (2.8), and

ii) indirect vaccination effect, i.e., $N\Delta_I = 76{,}300$ infections were averted in the unvaccinated population, using Eq. (2.9).

The ratio of indirect to direct vaccination effects is thus $(N\Delta_I)/(N\Delta_D) = 4.9$. Also, Eq. (2.16) gives an excellent approximation for $\Delta_I/\Delta_D$ that may be found without any of the above calculations, namely: $\Delta_I/\Delta_D \simeq 1/(R_0 - 1) = 5$.

The above example is intriguing because it demonstrates that at the population level the indirect vaccination effect can have a significant impact and, in this case, averts a relatively large number of infections, here some five times that of the direct vaccination effect. But the ratio is slightly misleading in that the unvaccinated pool where the indirect effects takes place is particularly large (450,000), in fact nine times larger than the relatively small vaccinated pool (50,000). To clearly understand this phenomenon, it essential to examine the effects on a "per capita" basis:

i) Direct vaccination effects are $(N\Delta_D)/(Nv) = 15{,}700/50{,}000 = 0.314$ infections averted *per* vaccinated individual.

ii) Indirect vaccination effects are $(N\Delta_I)/[N(1-v)] = 76{,}300 / 450{,}000 = 0.169$ infections averted *per* unvaccinated individual.

The ratio of indirect vaccination effect per unvaccinated individual to direct vaccination effect per vaccinated individual is the per-capita ratio (PCR):

$$PCR = \frac{N\,\Delta_I}{[(1-v)\,N]} \Big/ \frac{N\,\Delta_D}{v\,N} = \frac{v}{(1-v)} \cdot \frac{\Delta_I}{\Delta_D}. \qquad (2.17)$$

Thus, even though the indirect vaccination effect can be very large compared with direct vaccination effect (here five times larger), the per capita ratio of indirect to direct vaccination effect is $PCR = 0.54$. This measurement clearly shows that a vaccinated individual will still get much more benefit than an unvaccinated individual from vaccination, as might be expected.

## 2.4. The "per capita" effect

In summary, the Per Capita Ratio (PCR) of indirect to direct vaccination effects is defined as in Eq. (2.16), which combined with Eq. (2.11) gives:

$$PCR = \frac{v}{1-v}\frac{\Delta_I}{\Delta_D} = \frac{v}{1-v}\left(\frac{1-v}{v} - \frac{Z}{v\,Z^*}\right) \le 1. \qquad (2.18)$$

Figure 3c depicts PCR as a function of $R_0$ for a range of vaccination levels. It is important to note that, for a vaccine having 100% efficacy, all infections occur exclusively in unvaccinated individuals, so that indirect vaccination effects only impact the unvaccinated population.

As it is seen in Fig. 3c, the per capita PCR is always between 0 and 1. (PCR equals unity when the vaccination level equals or exceeds the herd immunity threshold $v_h$, and below the herd immunity threshold, the per capita ratio PCR is always smaller than 1.) Thus PCR indicates that the protection for unvaccinated individuals through the indirect vaccination effect cannot be larger than the direct vaccination effect for vaccinated individuals on a per capita basis. However, comparing this with the data from Fig. 3b, in the scenario where $1 < R_0 < 2$, we observe that $\Delta_I/\Delta_D > 1$, implying that the indirect vaccination effect consistently dominates the direct vaccination effect. As a result, solely focusing on the ratio $\Delta_I/\Delta_D$ can potentially exaggerate the perceived role of indirect vaccination effects, especially when the unvaccinated population is very large. Furthermore, it can be seen in Fig. 3c that given a particular $R_0$, a higher $v$ corresponds to a higher per capita ratio (PCR), suggesting that unvaccinated individuals can benefit more from a higher vaccination coverage. In contrast, as it can be seen in Fig. 3b or deduced from the approximation given in Eq. (2.16), the indirect- to direct-effect ratio $\Delta_I/\Delta_D$ is almost independent of the vaccination level. This clearly shows how $\Delta_I/\Delta_D$ can lead to misconceptions and highlights the importance of PCR for analysis of the effects of vaccination.

## 3. Shielding model of epidemics

In 2020, the SARS-CoV-2 virus emerged and led to a devastating global pandemic over the next years. Initially, and for at least a year, no vaccination was available for protection—it appeared the entire world population was susceptible. In the absence of vaccination, Weitz et al. [5] devised a mitigation strategy based on the fact that recovered individuals will always gain (at least short-term) immunity and can be used as "shields" to limit SARS-CoV-2 transmission. We will examine different shielding approaches and also discuss the role of indirect effects.

### 3.1. Shielding through recovered individuals

In practice, the strategy attempts to place susceptible members of the population as close as possible to anyone who has recovered from the disease. In other words, a susceptible individual will then tend to come into contact with recovered individuals more than other susceptible or infected individuals. In theory, the recovered individual will act to block transmission and reduce the number of chains of infection that could potentially reach the susceptible. The epidemic Final Size would also be expected to reduce. By relocating individuals and their contacts, the population mixing becomes non-random and thus more difficult to model. Nevertheless, Weitz et al. [5] devised a simple model to approximate the "shielding" effect. Their model assumes that there is a relative preference of $1 + \alpha$ that a given individual will interact with a recovered individual in what would be otherwise an interaction with a random individual. This type of interaction substitution is equivalent to assuming an effective contact rate ratio of $1 + \alpha$ for recovered individuals relative to the rest of the population. The larger is $\alpha$, the more chance that a susceptible individual is in the vicinity of a recovered, and therefore the larger is the shielding effect.

The following modified SIR equations [5] involve shielding into the process:

$$\begin{aligned}\frac{dS}{dt} &= -\beta\frac{SI}{1+\alpha R},\\ \frac{dI}{dt} &= \beta\frac{SI}{1+\alpha R} - \gamma I,\\ \frac{dR}{dt} &= \gamma I,\end{aligned} \quad (3.1)$$

where $S, I, R$ are the fractions of susceptible, infectious, and recovered individuals, respectively, and $S(0) = 1, I(0) = R(0) = 0$. Given that $S + I + R = 1$, the term $1 + \alpha R$ can be thought of as $S + I + (1 + \alpha)R$. This assumes an effective contact ratio of $1 + \alpha$ for a recovered individual relative to the rest of the population $(S + I)$. The transmission rate is given by the term $\beta/[1 + \alpha R(t)]$, and thereby clearly, the more recovered individuals there are, or the higher is $\alpha$, the lower will be the epidemic transmission through the population. The time-varying basic reproduction number is defined as $R_0(t) = \beta/[(1 + \alpha R(t)) \cdot \gamma]$, where $\beta$ is the infection rate, and $\gamma$ is the recovery rate.

Through numerical simulations Weitz et al. [5] show that shielding acts to reduce the epidemic peak and shortens the duration of epidemic spread. By directly solving the following Eqs. (3.4, 3.5), without relying on numerical simulations, we show here the effect of shielding on the Final Size of the epidemic as a function of the basic reproduction number $R_0 = \beta/\gamma$ in Fig. 4a, and as a function of shielding strength $\alpha$ in Fig. 4b. It is clear from the results that very large levels of shielding are required to make a substantial impact on the size of the epidemic. An important observation is that any shielding, no matter how strong (even for $\alpha = 20$), slows down the epidemic but never stops it completely (i.e., via crossing the herd immunity threshold). Intuitively, this might be expected, since in situations when there are few recovered individual available to provide protection, shielding becomes less effective. In other words, shielding via "recovered" population has a large effect when the epidemic is largely spreading, and the more shielding is applied, slowing down the epidemic results in a decrease in the shielding effect itself; see Fig. 4b where increasing shielding strength asymptotically loses effect.

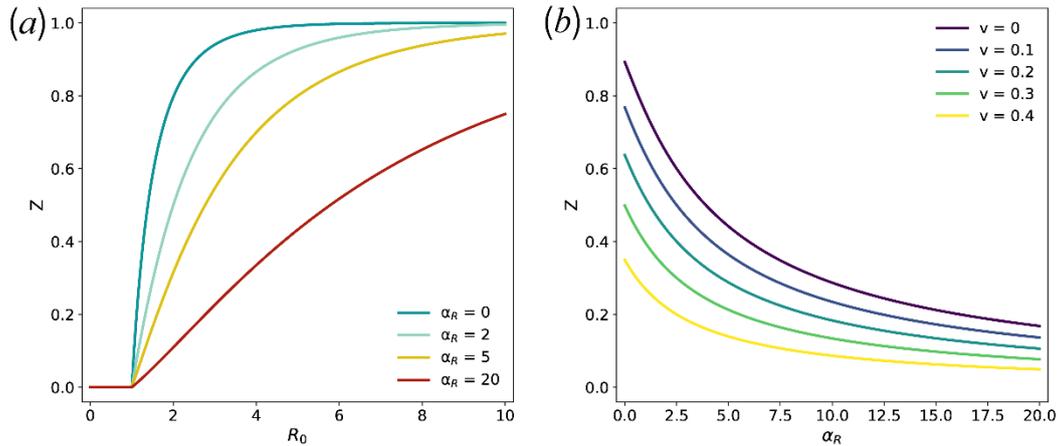

**Figure 4.** (*a*) The Final size of the epidemic $Z$ as a function of the basic reproduction number $R_0$ for initial vaccination coverage $v = 0$ with shielding strength $\alpha = \alpha_R = 0, 2, 5, 20$. (*b*) The Final size of the epidemic $Z$ as a function of the shielding strength $\alpha$ (or $\alpha_R$, since shielding is implemented on Recovered individuals only) for $R_0 = 2.5$ separately for different vaccination levels $v$; calculated by solving Eqs. (3.4-3.5).

### 3.2. Shielding through both vaccinated and recovered individuals

Here we extend the shielding strategy [5] to a more general mitigation strategy where some (even limited) vaccination is available and thus both recovered and/or vaccinated individuals can be used as shields. In the *extended shielding model*, a susceptible individual has a relative preference of $1 + \alpha_V$ for contact with a vaccinated and $1 + \alpha_R$ with a recovered individual relative to other individuals. The new model can be formally presented with the following equations:

$$\frac{dS}{dt} = -\beta \frac{SI}{1 + \alpha_V v + \alpha_R R},$$
$$\frac{dI}{dt} = \beta \frac{SI}{1 + \alpha_V v + \alpha_R R} - \gamma I, \quad (3.2)$$
$$\frac{dR}{dt} = \gamma I,$$
$$\frac{dN}{dt} = \alpha_R \gamma I,$$

where $S, I, R$ is the fraction of susceptible, infectious, and recovered individuals, respectively, and we define the function $N(t) = 1 + \alpha_V v + \alpha_R R(t)$ which plays a key role in our analysis of the model. The fraction of vaccinated individuals is set as a constant $v$, and thus $S(0) = 1 - v$, $I(0) = R(0) = 0$. Similar to the base shielding model (Eq. (3.1)), given $S + I + R + v = 1$, the term $1 + \alpha_V v + \alpha_R R$ can be thought of as $S + I + (1 + \alpha_V)v + (1 + \alpha_R)R$. Then, the transmission rate is given by the term $\beta/[1 + \alpha_V v + \alpha_R R(t)]$, and thus, the time-varying basic reproduction number is given by $R_0(t) = \beta/[(1 + \alpha_V v + \alpha_R R(t))\gamma]$, where $\beta$ is the infection rate, and $\gamma$ is the recovery rate.

### 3.3. Analysis of the extended shielding model

The extended shielding model (Eq. (3.2)) can be investigated analytically to characterize the effect of shielding when both vaccinated and recovered population are participating. Here, we examine the relationship between the Final Size of the epidemic ($Z$) and the shielding strength $\alpha_V$ and $\alpha_R$.

**Shielding via vaccinees only ($\alpha_V > 0, \alpha_R = 0$).** This is the case where recovered individuals are not used as shields, and the shielding is based entirely on vaccinated individuals. As before, we assume that at $t = 0$ initial conditions are $S(0) = 1 - v$, $I(0) = R(0) = 0$. Since $\alpha_R = 0$, the system in Eq. (3.2) is equivalent to a standard SIR epidemic except now with $R_0(v) = \beta/[(1 + \alpha_V v)\gamma]$. Thus, the effective reproduction number is given by $R_{\text{eff}}(t) = R_0(v) S(t)$, and an epidemic will occur at $t = 0$ only if $R_{\text{eff}}(0) > 1$, so that $I(t)$ would have initial positive growth. The condition for epidemic break-out can be worked out for the vaccinated shielding strength as:
$$\alpha_V < [\beta(1-v)/\gamma - 1]/v. \quad (3.3)$$
The Final Size of the epidemic $Z(v)$ is as before defined by Eq. (2.4). Thus, increasing vaccinated-shielding ($\alpha_V$), decreases $R_0$, and this in turn decreases the Final Size $Z$ until it reaches zero when $\alpha_V = [\beta(1-v)/\gamma - 1]/v$. This means that with a large enough vaccinated-shielding, an epidemic can be prevented.

**Shielding via both vaccinees and recovered individuals ($\alpha_V, \alpha_R > 0$).** The equilibrium of Eq. (3.2) occurs when all time-derivatives are set to zero. Note that at equilibrium we must have $\lim_{t\to\infty} I(t) = I^* = 0$. Suppose also that at equilibrium $\lim_{t\to\infty} N(t) = N^*$ and $\lim_{t\to\infty} R(t) = R^*$. Since the Final Size of the epidemic $Z$ is exactly $R^*$, and $N^* = 1 + \alpha_V v + \alpha_R R^*$, thus $Z$ can be derived as:
$$Z = (N^* - 1 - \alpha_V v)/\alpha_R. \quad (3.4)$$

Our analysis of the extended shielding model (Eq. 3.2) in Supplementary Note 4 reveals that $N^*$ satisfies the equation below:
$$N^{*c+1} - (\alpha_R(1-v) + 1 + \alpha_V v)N^{*c} + \alpha_R(1-v)(1 + \alpha_V v)^c = 0, \quad (3.5)$$

where $c = \beta/\alpha_R \gamma$ and $1 + \alpha_V v \leq N^* \leq 1 + \alpha_V v + \alpha_R(1-v)$. In Supplementary Note 4 we show in detail that there is always only one root of Eq. (3.5) that satisfies the range of $N^*$. Using numerical procedures, Eq. (3.5) can be solved to obtain the unique value of $N^*$, and substituting $N^*$ in Eq. (3.4), gives the Final Size of the epidemic $Z$. For this scenario, the time-varying basic reproduction number is given by $R_0(t) = \beta/[(1 + \alpha_V v + \alpha_R R(t))\gamma]$, the effective reproduction number is given by $R_{\text{eff}}(t) = R_0(t) S(t)$, and an epidemic will occur only if $R_{\text{eff}}(t) > 1$ at $t = 0$, i.e., $\alpha_V < [\beta(1-v)/\gamma - 1]/v$. (Note that the condition for epidemic break out is the same as Eq. (3.3) calculated for the case where $\alpha_R = 0$, investigated earlier.)

The effect of shielding via vaccinated population is demonstrated by the results in Fig. 5a, obtained by applying Eqs. (3.4) and (3.5) derived for the extended shielding model formulation (Eq. (3.2)). The solid curves represent the case where there is no recovered shielding ($\alpha_R = 0$) and dashed curves represent the case where recovered- and vaccinated shielding increase together ($\alpha_R = \alpha_V$). As one can expect, the larger is $\alpha_V$, the smaller is the Final Size of the epidemic. Clearly, for different vaccination levels color-coded in Fig. 5a, the Final Size of the epidemic reaches zero only when vaccinated shielding strength crosses the herd immunity threshold $\alpha_V = [\beta(1-v)/\gamma - 1]/v$ (see Eq. (3.3)), regardless of the recovered shielding strength $\alpha_R$. Moreover, the results in Fig. 5a show that for a higher initial vaccination coverage $v$, the herd immunity threshold for vaccinated shielding strength $\alpha_V$ is lower (consistent with Eq. (3.3)). In other words, the higher is the vaccination level, the less vaccinated shielding is required to achieve herd immunity. Interestingly, adding recovered-shielding further reduces the Final Size between $0 \leq \alpha_V < [\beta(1-v)/\gamma - 1]/v$, but only increasing $\alpha_V$ is able to make $R_{\text{eff}}(0) \leq 1$ and completely curb the epidemic. This is seen by solid and dashed curves of each particular color (corresponding to a particular vaccination level) reaching $Z = 0$ at the same $\alpha_V$ threshold independent of $\alpha_R$. In Supplementary Note 4, we use numerical simulations to find the relationship between the Final Size of the epidemic and $\alpha_V, \alpha_R$, and obtain identical results that confirm those shown in Fig. 5a.

It is intuitive that using vaccinated individuals as shields can increase the indirect vaccination effect, thereby increasing the PCR. It is thus helpful to use PCR to explain the curbing effect of vaccinated-shielding strength $\alpha_V$ on epidemics. In Fig. 5b, we show the effect of shielding on the PCR as a function of shielding strength $\alpha_V$, for different levels of vaccination $v$ in the initial population. As it is seen in Fig. 5b, increasing vaccinated-shielding strength $\alpha_V$ directly increases PCR, with higher vaccination level strengthening this effect. Also the results confirm the epidemic break-out (or herd-immunity if viewed inversely) condition in Eq. (3.3), with PCR reaching unity exactly at $\alpha_V = [\beta(1-v)/\gamma - 1]/v$ (marked by the dashed lines associated with different vaccination levels).

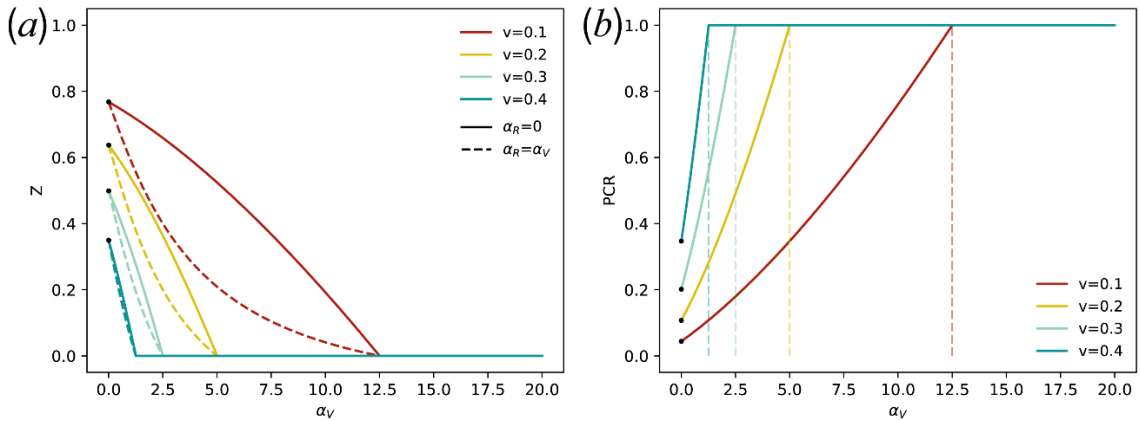

**Figure 5.** (*a*) The Final size of the epidemic $Z$ as a function of the shielding strength $\alpha_V$ for different vaccination levels $v = 0, 0.1, 0.2, 0.3, 0.4$, separately demonstrated for no recovered-shielding $\alpha_R = 0$ (solid curves), and increasing recovered-shielding $\alpha_R = \alpha_V$ (dashed curves). (*b*) PCR as a function of vaccinated-shielding strength $\alpha_V$ for different vaccination levels $v = 0, 0.1, 0.2, 0.3, 0.4$. The results in both panels are calculated based on the PCR definition in Eq. (2.18) and the solution of Eqs. (3.4) and (3.5); all data are generated with assumed $R_0 = 2.5$.

## 4. Discussion and conclusion

Unlike the approaches used in previous studies, such as the equilibrium analysis of Eichner et al. [4] and the approximate discrete-time SIR model of Scutt et al. [11], this paper adopts the Final Size formula to explore the impact of vaccination. This method allows us to obtain many useful results all analytically and unpack vaccination effects and mitigation strategies, and also paves the way for advanced analytical evaluation of current and future strategies. Our results indicate that the ratio of indirect to direct vaccination peaks at the herd immunity threshold $v_h = 1/R_0 - 1$ (Fig. 3a), with the ratio being substantial for outbreaks with lower vaccination coverage $v$ (Figs. 3a and 3b). Although the indirect vaccination effect can be substantial—potentially many times greater than the direct vaccination effect— the presence of a large unvaccinated population can change how this should be interpreted. The influence of the size of unvaccinated population is reflected better in per capita ratio of indirect- to direct-effect of vaccination (which we denoted PCR), and can be an important tool to analyze vaccination strategies as shown by analytical results in this manuscript. As expected, the PCR measure reveals that the benefits of the indirect vaccination effect for an unvaccinated individual can never surpass the direct protection of vaccine for a vaccinated individual.

Either limiting the transmission of the infectious disease or reducing the number of susceptible individuals (e.g., through vaccination) can effectively reduce the Final Size of the epidemic. While the strategy of using recovered individuals as shields to limit the transmission of SARS-CoV-2 has been widely discussed [17-19], it has the disadvantage of requiring large-scale serologic testing, making it virtually impractical [5]. With regards to vaccination, achieving an initial coverage beyond the herd immunity threshold can reduce the Final Size of the epidemic to zero, but that usually requires vaccinating a substantial proportion of a population. Moreover, vaccine availability may be limited, especially in less developed countries. Our proposed extension of the shielding model can be an effective response. It suggests using vaccinated individuals as shields without the help of, or in addition to, the shielding of the recovered population. This model achieves the goal of reducing the Final Size of the epidemic to zero for any non-zero vaccination coverage, given that the shielding strength via vaccinated population $\alpha_V$ is sufficiently large, i.e., $\alpha_V \geq [\beta(1 - v)/\gamma - 1]/v$. Tracking vaccinated individuals is more feasible than tracking recovered individuals, making the extended shielding a more practical mitigation strategy, assuming a vaccine is available.

It is important to mention that our study primarily focused on epidemics with fixed $R_0$ and initial vaccination coverage $v$, assuming a 100% vaccination efficacy. In more realistic scenarios, these parameters change over time instead of remaining constant. Therefore, in certain extreme cases, our main conclusions regarding indirect and direct vaccination effects may not be applicable. For instance, if no one is vaccinated before the epidemic and the rate of vaccination during the epidemic is slow, the epidemic may conclude before a significant fraction of the population is vaccinated. Additionally, the rate of immunity loss from infection or vaccination will also impact the results [9, 20, 21]. These complications suggest important future research directions.

## Funding

LS was supported by Australian Research Council (Australia) (DP240102585). The funders had no role in study design, data collection and analysis, decision to publish, or preparation of the manuscript.

## Contributions

LS and LL conceived the work, LL, LS, HH – programming, calculations, drafting the paper. LS, LL, HH and RS reviewed the work critically. All authors approved the submission.

## Conflict of interest declaration

The authors declare that they have no competing interests.

## Acknowledgements

None.